# MINERALOGICAL CHARACTERISTICS OF SPECIMENS OF A METEORWRONG "FALL" FROM NW IRAN.

H. Pourkhorsandi and H. Mirnejad, Department of Geology, Faculty of Science, University of Tehran, Tehran 14155-64155, Iran (hkhorsandi@khayam.ut.ac.ir; mirnejad@khayam.ut.ac.ir)

**Introduction:** In the early hours of October 22, 2011, people of a small town in NW Iran, called Khameneh (38°11'47" N, 45°38'14" E) noticed sounds of some colliding objects to home roofs and yard floors. News of the incident propagated in the media and internet immediately and a number of hazelnut-sized specimens were collected by local people. A field trip to the area by first author led to collection of three samples, weighing between 6 and 11 grams. These samples were, however, scattered in an area of ~200m in diameter that was very small compared with strewn fields of meteoritic falls [e.g. 1].

**Macroscopic investigations:** The presence of light and dark regions on the surface of collected samples gives them an overall white-black appearance (Fig. 1). Preliminary investigations showed that the rocks had densities of ~ 2.2 g/cm$^3$. In a cut surface, the interior is darker than surface and consists of abundant vesicles (Fig. 2). This porosity justifies the low density of samples.

**Microscopic investigations:** Two thin sections were prepared for preliminary microscopic examinations. Fine grained, rounded and brownish minerals with low birefringence are the most dominant constitutes. Minerals are very similar to phylosilicates in appearance and in some cases have been aggregated in a pseudo-olivine structure (Fig. 3). Droplet chondrule – like aggregates are also present.

**Mineralogy:** For detailed mineralogical investigation of the specimens, SEM and mineral chemistry examinations were undertaken at Origins Laboratory, Department of the Geophysical Sciences, University of Chicago (SEM image: Fig. 4).

Three primary solid phases were identified. Larnite or bredigite ($Ca_2SiO_4$) is one of them that is a rare and little known compound. Larnite ($\beta$-$Ca_2SiO_4$) forms part-along with forsterite ($Mg_2SiO_4$), fayalite ($Fe_2SiO_4$), and thephroit ($Mn_2SiO_4$) of the monticellite and knebelite series (general group of olivine) [2]. However, despite its scarcity, larnite is found in different natural settings [3,4], almost always under thermodynamic conditions of around 0.2–1 kbar and 1000–1100°C. This mineral can also be formed artificially, especially during the synthesis of high technology refractory and ceramic materials, and as a mineral component of some industrial slags and portland cements [5]. The other phase is hatrurite ($Ca_3SiO_5$), a mineral similar to larnite that is often a paragenesis of larnite in metamorphic suites. Alite is a name that is given for the industrial equlivalet of harturite. Brownmillerite ($Ca_2AlFeO_5$) is the other primary mineral phase that is similar in formation conditions to previous minerals.

Field works in the region failed to find any evidence of mentioned minerals in metamorphic rock suites and in ceramic - brick factories around Khameneh. For investigation of any possible relationship between Khameneh specimens with artificial compoounds formed in a Portlant Cement Company (since the primary materials for producing cement is limestone (Ca rich) and clay (Al, Si rich)) at about 40 km NE of the town, a visit to the furnace of the company carried out and it was surprising to find out that furnace specimens were very similar in appearance to those of "had fallen" at Khameneh. Furnace rocks, named as Portland Cement Clinkers, are small, dark grey nodular materials made by heating ground limestone and clay at a temperature of about 1400-1500°C [6]. The nodules are grinded up to fine powders to produce cement. Nodules range in size from 1mm to 25mm or more and are composed mainly of calcium silicates, typically 70%-80%. Their microscopic textures are very similar to the textures of Khameneh stones. Clinkers are mainly composed of alite ($Ca_3SiO_5$ = Hatrurite), belite ($Ca_2SiO_4$ = larnite, bredigite), and also brownmillerite [7] (Fig. 5).

Khameneh event is somehow very similar to an event that occurred on 21$^{st}$ June, 1994, in the city of Getafe (Spain) [8]. A moving car and its driver were struck by a falling rock. Eighty-one additional fragments were later recoverd, which all pointed toward a meteorite fall. A study of the composition of this object revealed an ultrarefractory material that is very similar to mineral composition of Khameneh specimens (Fig. 4). Based on physical characteristics of "fall" and collected fragments and mineral chemistry of specimens, we concluded that Khameneh specimens were indeed man-made objects (portland cement clinkers) and thus Khameneh event was not a meteorite fall-related phenomenon.

**Acknowledgments:** The authors would like to thank Dr. A. Pourmand (RSMAS) and Origins Laboratory research team for mineral chemistry analyses.

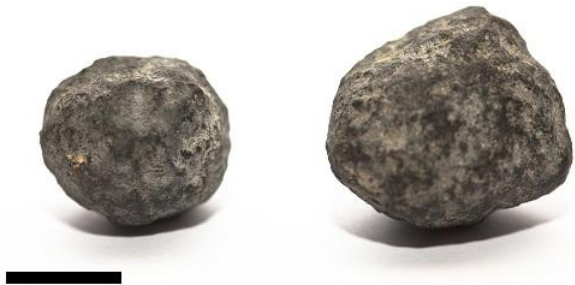

*Fig. 1. Two fragments of Khameneh alleged meteorites weighing 6 and 11 grams, on the left and right side, respetively. Scale bar: 10mm*

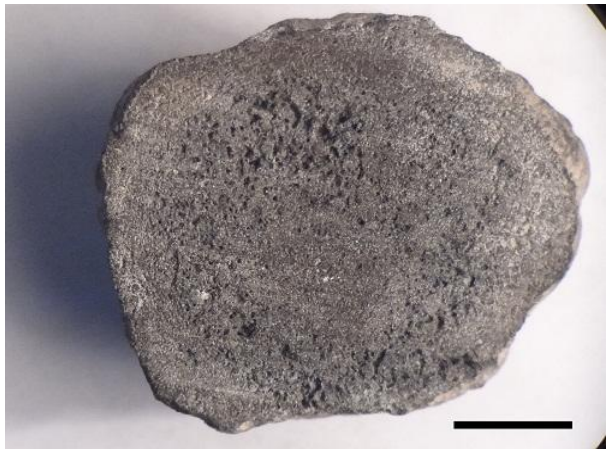

*Fig. 2. Interior of the specimens are highly porous. (note the low porosity at rim). Scale bar: 5mm*

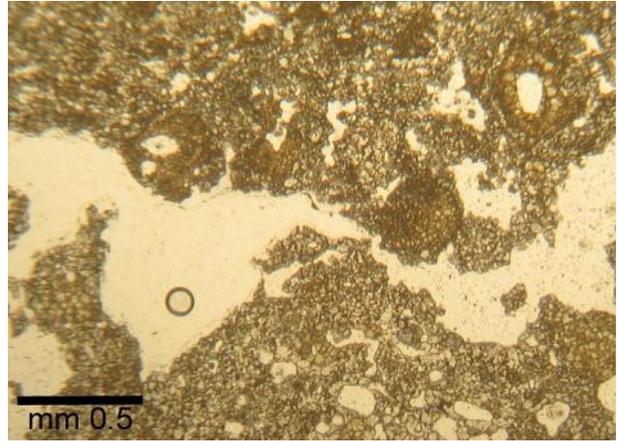

*Fig. 3. A microphotograph of Khameneh specimen.*

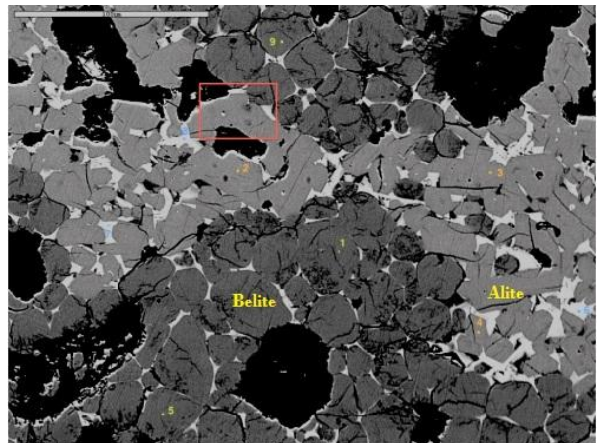

*Fig. 4. SEM image shows three primary solid phases. Dark grey phase is $Ca_2SiO_4$ (belite), light grey is $Ca_3SiO_5$ (alite) and white interestital part is brownmillerite. Scale bar (upper left): 100μm*

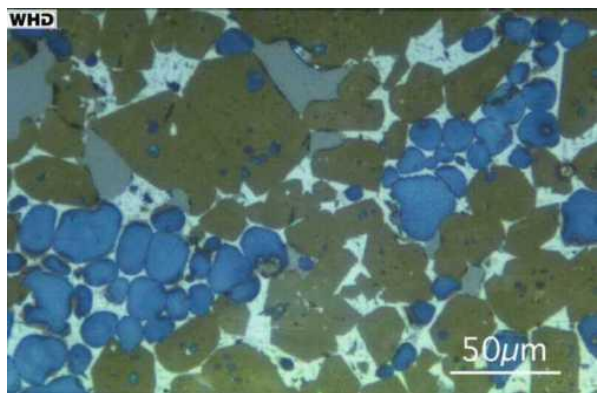

*Fig. 5. Optical microscope image (polished section) of a clinker nodule. Brown crystals are alite, blue crystals are belite, bright interstitial material is mainly ferrite, with small dark inclusions of aluminate. (www.understanding-cement.com)*